\documentstyle[proceedings,graphics,bibstyle]{crckapb}

\def\spose#1{\hbox to 0pt{#1\hss}}
\def\lta{\mathrel{\spose{\lower 3pt\hbox{$\mathchar"218$}}
     \raise 2.0pt\hbox{$\mathchar"13C$}}}

\begin{opening}
\title{SECULAR EVOLUTION OF GALAXY MORPHOLOGIES}
\author{D. PFENNIGER, L. MARTINET}
\institute{Geneva Observatory, University of Geneva\\
           CH-1290 Sauverny, Switzerland}
\author{F. COMBES}
\institute{DEMIRM, Observatoire de Paris\\ 
	61 Av. de l'Observatoire, F--75014 Paris, France}
\end{opening}

\runningtitle{EVOLUTION OF GALAXY MORPHOLOGIES}

\begin{document}

\begin{abstract}
Today we have numerous evidences that spirals evolve {\it
dynamically\/} through various secular or episodic processes, such as
bar formation and destruction, bulge growth and mergers, sometimes
over much shorter periods than the standard galaxy age of $10-15$ Gyr.
This, coupled to the known properties of the Hubble sequence, leads to
a unique sense of evolution: from Sm to Sa.  Linking this to the known
mass components provides new indications on the nature of dark matter
in galaxies.  The existence of large amounts of yet undetected dark
gas appears as the most natural option.  Bounds on the amount of dark
stars can be given since their formation is mostly irreversible and
requires obviously a same amount of gas.
\end{abstract}

\section{Introduction}
Until the recent years the concepts prevailing in understanding galaxy
evolution have been largely dominated by the ELS scenario (Eggen,
Lynden-Bell \& Sandage 1962).  The views issued from it favour a {\it
synchronous\/} and rapid ($\approx 100\,\mathrm{Myr}$) formation of
the galaxies at a particular early time in the universal expansion.
In this context, the properties of present galaxies, such as their
typical scale and mass, must depend directly on the initial conditions
fixed by the physical state of the Universe at the ``galaxy formation
epoch''.  The only subsequent significant evolution in galaxies to be
discussed was the slow changes in their stellar populations.  Dynamical
processes could be assumed to be settled, and therefore ignored.

Since then major observational and theoretical progresses of direct
relevance occurred, which resulted in a gradual shift in the meaning
of galaxy evolution.  It suffices to remind that basically all the
advances in non-linear dynamics, including the recognition of the
fundamental r\^ole of chaos, and in computer simulations of galaxies
have been made in between.

Today, in all its steps the ELS scenario is no longer tenable, as
explained in the following.  This is still not perceived as a
necessity in fields loosely connected with the recent advances in
gravitational dynamics.  
 
An instance of inadequation between the ELS scenario and more recent
works appears in cosmological simulations at several
$100\,\mathrm{Mpc}$ scale (see e.g., White 1994).  In such simulations
nothing like homogeneous collapses do occur, instead hierarchical
clustering proceeds at all the computable scales with different
speeds.  This implies that galaxies and galaxy clusters, exactly as
stars, form in different regions of the sky {\it asynchronously}.  The
formation process covers several dex of time-scales, so not only the
free-fall time, but merging and later infall participate to it.  For
many galaxies the formation/evolution should be considered as not
terminated even now. The galaxy age looses its original meaning
because the aging, traced by various observables, may occur at widely
different speeds.

A central aspect not considered in the ELS picture is the likely
coupling of dynamics with stellar activity.  In the recent years the
large FIR emission of spirals, which even largely dominates the light
in starburst galaxies, was found substantial particularly over the
``late'' part of the spiral sequence.  The FIR emission, coming mostly
from UV and visible light absorbed and recycled in the FIR by dust, is
consistent with the also recent recognition of the partial opacity of
the optical region of spirals (e.g. Davis et al.~1993).  It turns out
that the total bolometric luminosity is comparable to the power that
dynamics can exchange (Pfenniger 1991b).  This coincidence is best
explained by a coupling of star formation and dynamics via a feed-back
mechanism (Quirk 1972; Kennicutt 1989).  The interesting aspect of
this coupling is that the systematic global properties of galaxies are
then no longer necessarily determined by the initial conditions of
collapse.  As for stars, the galaxy properties would then be encoded
in their internal small scale physics, i.e.~star formation and ISM
physics.  This may solve the old problem of the absence of galactic
scale at the radiation-matter decoupling epoch, particularly if galaxy 
formation covers a sizable fraction of the Hubble time.  The galactic
scale would result mainly from the balancing of stellar activity
effects, particularly during starburst phases, and dynamics.

The Hubble sequence is most probably an incomplete sample.  For
example many low surface brightness galaxies may well be missed
(e.g.~Bothun et al.~1990).  However, whenever fast morphological
changes do occur in normal galaxies (bars, mergers), they must
correspond to shifts {\it within the sequence}, because the already
existing stars certainly survive the changes.

The general properties of the Hubble sequence have been progressively
determined (e.g.~de Vaucouleurs 1959; Broeils 1992; Roberts \& Haynes
1994; Zaritsky et al.~1994).  In order to extract useful information
of this ``zoo'', we must consider only the most general properties,
keeping in mind that galaxies form a variety of objects with different
ages and aging speeds.  For example mergers (Schweizer 1993) certainly
accelerate morphological changes at speeds which depend on the
fortuitous interaction strengths.

The following list summarises the main trends and properties along the
spiral sequence {\bf from Sm to Sa}, that are useful for the discussion.

\def\Up{$\nearrow$}
\def\Down{$\searrow$}

\begin{tabular}{lll} 
\hline\noalign{\smallskip}
Total mass&\Up&$M_{\mathrm{tot}} \approx 1\to100 \cdot 10^{10}\,\mathrm{M_{\odot}}$\\
Kinetic energy&\Up&$\frac{1}{2} V_{\mathrm{max}}^2 \approx 25 \to 500 \,\mathrm{eV/nucleon}$\\
Bulge-disk ratio&\Up&$ L_{\mathrm{sph}}/L_{\mathrm{tot}}\approx 0\to0.6$\\
Symmetry&\Up&\\
Metallicity&\Up&$12+\log(\mathrm{ O/H}) \approx 8.3\to9 $ \\
Detected gas&\Down&$\displaystyle\frac{M_{\mathrm{HI+H_2}}}{M_{\mathrm{tot}}}\approx 0.10\to0.07$, 
$\displaystyle\frac{M_{\mathrm{HI+H_2}}}{M_{\mathrm{stars}}}\approx1.4\to 
	0.1$ \\
Dark matter&\Down&$M_{\mathrm{dark}}/M_{\mathrm{lum}} \approx 10 \to 1$ \\ 
\noalign{\smallskip}\hline\noalign{\vskip-1mm plus 1mm minus 1mm}
\end{tabular}

\section{Sense of Evolution from Irreversible Processes}
In fact, already a systematic sense of evolution is clear by making a
list of the major irreversible processes known in spirals.  Each of
them gives a possible criterion and a sense of aging.

1) The energy dissipation in gravitating systems is measured by the
present amount of kinetic energy, which equals the minimum energy the
system had to release in order to reach the present bound state.  In
spirals the rotation speed is an excellent indicator of the specific
energy dissipated since the rotation curves vary slowly with radius.
Because disks are systems having lost the maximum of energy while
conserving angular momentum well, further energy dissipation
necessarily implies dissipation of angular momentum, which is best
achieved by some mass transport and breaking of axisymmetry.  Bars and
spiral arms are just manifestations of this necessity.  The energy
factor already indicates clearly that the Sa side of the spiral
sequence is energetically more evolved than the Sm side.

2) Building a central bulge or spheroid by heating a disk, by whatever
process (see next Sect.), is also a stellar dynamical irreversible
process, because there is no way to ``cool'' stars following fat
orbits back toward circular orbits.  From the stellar dynamical point
of view Sa galaxies with big bulges are thus more evolved than
bulgeless Sm galaxies.

3) Overall, if galaxian shapes tend toward some attractors, the degree
of organisation and symmetry toward these shapes is a sign of
evolution.  Clearly the spiral sequence looks increasingly organised
in the sense Sm to Sa, which is also consistent with the decreasing
spiral pitch angle.

4) The transformation of gas into stars in cold clouds is mainly an
irreversible process, star formation locking most of the mass for
time-scales longer than galaxy ages.  So the ratio of stellar to gas
mass is a tracer of evolution, and Sa's are more evolved in this
respect than Sm's.  In this context, the dark matter fraction
decreasing systematically along the spiral sequence to low
``non-problematic'' values on the Sa side is remarkable.

5) Obviously related to the previous criterion, the nucleo\-synthesis
within stars is also irreversible, the more metal rich and dusty
galaxies have a longer history through the internal activity of their
stars.  Sa's are more enriched than Sm's, which again indicates a
sense of evolution.

Finally, the total mass along the sequence increases strongly from Sm
to Sa in average, though with a considerable spread at constant type.
This is not astonishing because the classification criteria are mass
independent.  We interpret the mass trend as indicative that massive
proto-galaxies evolve in average faster than light ones.  The
proto-galaxies of today (Im's, Sm's) are probably lighter than the
corresponding ones in the past.

In summary, the only consistent sense of aging along the spiral
sequence is from Sm to S0. The important consequence is that
proto-galaxies would then be mostly bulgeless gas rich disks like
Sm-Sd's, or even pure gas disks, instead of initial spheroid dominated
galaxies of the ELS scenario.

\section{Evolution from Dynamics and Observations}
Contemporary to the ELS scenario, Safronov (1960) and Toomre (1964)
realised the unexpected fact that gaseous and stellar disks with too
much circular motion are gravitationally unstable. So energy
dissipation with angular momentum conservation brings first collapsing
gravitating systems toward disk shapes with an increasing fraction of
kinetic energy in rotational motion.  But subsequent dissipation
brings disks ineluctably toward a global instability.  A first ground
was found that disk galaxies may be dynamically unstable, so may
evolve with dynamical time-scales.

Shortly hereafter computer simulations of stellar disks (e.g.  Miller
\& Prendergast 1968; Hohl \& Hockney 1969) allowed to simulate the
non-linear phase of disk instability.  They showed a systematic
tendency to produce a robust bar.  These results illustrated an
example of major and fast morphology change (within a couple of
rotational periods) of galaxy type from non-barred to
barred.

The other significant proposition in the 70's came from Toomre \&
Toomre (1972) in which ellipticals may result from the merging of
spirals or other ellipticals.  Another case of major and fast change
of galaxian morphology was put forward.  Despite many resistances this
scenario appears today as the most natural way of forming ellipticals,
although it is not necessarily the only one.  In fact, the more
violently a galaxy is shaken, for whatever reason, the more it ends up
like an elliptical.

In the 80's the bar phenomenon was investigated in more depth.  From
observational material, Kormendy (1982) pushed forward the idea of
secular evolution in barred galaxies.  The reason why bars do exist in
the first place was understood by studies of their periodic orbits in
the plane (Contopoulos 1980, Athanassoula et al.~1983) and in 3D
(Pfenniger 1984, 1985).  It was discovered that bars may evolve into
boxy bulges via vertical resonances boosting bending instabilities
transverse to the plane (Combes \& Sanders 1981; Combes et al.~1990;
Raha et al.~1990).  It became also clear that chaotic orbits are
playing an important role in bars.  Later it was understood how the
accretion of only a few percents of mass within the Inner Lindblad
Resonance (ILR), either by dissipation of gas (Hasan \& Norman 1990;
Pfenniger \& Norman 1990) or by dynamical friction on galaxy
satellites (Pfenniger 1991a), may rapidly destroy a bar into a
spheroidal component of similar size.  This led to an increased
confidence that many of the non-linear events and morphologies
observed in galaxies and in N-body simulations, can be interpreted,
and even predicted, via the knowledge of the underlying periodic
orbits (Pfenniger \& Friedli 1991).  These studies of bars showed that
isolated galaxies must also be seen as structures with possible fast
dynamical evolution phases.

In the recent years more works have completed the above picture.
Secondary bars (Friedli \& Martinet 1993), gaseous and star formation
effects (e.g.~Friedli \& Benz 1993, 1995), interactions with external
galaxies and mergers (e.g.~Barnes 1992) continue to be investigated.  In
any cases, these additional complications make even harder to freeze
galaxy morphologies beyond a few Gyr.  The obvious requirement is then
to understand the general time-sequence of galaxy morphologies.

Independently of dynamics, several observational results strongly
indicate significant secular evolution within much less than 10 Gyr:

1) From halo stellar cluster abundances Searle \& Zinn (1978) arrived
to an alternative scenario to ELS.  From the observations they
concluded that the Milky Way stars did form inside-out over several
Gyr.

2) In galaxy clusters the Butcher-Oemler (1978) effect (see also Rakos
\& Shombert 1995) indicates that galaxies are increasingly bluer at
higher redshifts.  Furthermore, the morphology-radius relationship
(Dressler 1980; Whitmore 1993) shows that the majority of galaxies at
cluster periphery are spirals, as in the field, but these are replaced
by lenticular and then ellipticals at smaller radii.  To a dynamicist
this relationship tells that spirals do not survive to a single center
crossing or so, because the galaxies within a cluster are expected to
move on rather elongated orbits.  If correct the spiral morphological
evolution accelerated by environmental disturbances is then directly
revealed.  Either spirals end in part as ellipticals, and/or they are
largely dissolved and contribute to the cluster hot gas and 
metallicity.  

\section{Constraints from an Evolutive Spiral Sequence}
If we take the published data about the ratios of the different known
matter forms along the spiral sequence (e.g.~Broeils 1992 for the
stars and HI; Young \& Knezek 1989 for H$_2$ derived from CO
emission), we can solve for the mass fraction of each component:
stars, HI, H$_2$ from CO, and the rest, dark matter (for more detail,
see Pfenniger 1996).  We obtain the composition diagram of the spiral
sequence shown in Fig.~1.

\begin{figure}
\includegraphics[15mm,64mm][125mm,126mm]{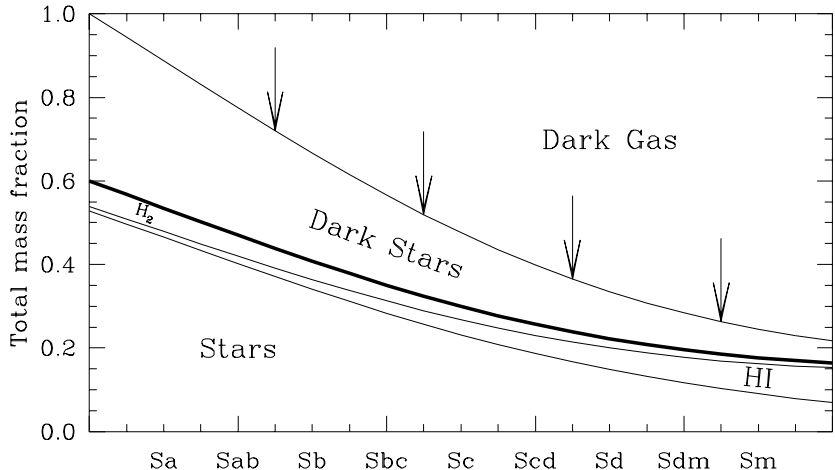}
\vspace{-2mm}
\caption{Composition diagram along the spiral sequence.  The thick
line separates the detected matter from the dark matter.  The dark
stellar-like objects (DS) are assumed here to be proportional to the
stars.  The arrows indicate that the DS mass is an upper limit and the
dark gas a lower limit.}
\vspace{-2mm}
\end{figure}

Now if spirals do evolve along the sequence from Sm to S0, then
several constraints follow (Pfenniger, Combes \& Martinet~1994;
Martinet 1995).  We must consider that stars or jupiters are made from
the gas and lock most of it for $\gg 5-10\,\mathrm{Gyr}$, and that for
dynamical reasons little accretion (less than a few \% in mass) can
occur transversally to a stellar disk without heating it to too high
values (T\'oth \& Ostriker 1992).  

Most of the usual dark matter candidates such as CDM particles,
neutrinos, jupiters, brown dwarfs, and black-holes, are inconsistent
with the spiral sequence properties including dynamical evolution,
mainly because evolutive phases (e.g. a merger) can be sporadic.
Since the star fraction increases from Sm to Sa in a proportion
exceeding the detected gas content, everything happens as if dark
matter is transformed into stars, i.e. a substantial fraction of dark
matter should be dark gas.  We have explained elsewhere (Pfenniger \&
Combes 1994, and elsewhere in this volume) the many reasons why today
we can consider this conservative candidate, for long totally
neglected, as worth to be more investigated.

Others did arrive also to the conclusion that much more gas should
exist in spirals.  Toomre (1981) argued that the cold dynamics and
chaotic structure of Sc's require much more gas than observed to
maintain their patchy structure.  Also just to sustain the present
star formation rates over several Gyr Sc's require much more gas
(Larson et al.~1980). 

Not all the dark matter in spiral is necessarily dark gas, because the
uncertainties on the stellar mass contributed by brown dwarfs and dark
remnants, or by stars obscured by dust, are still substantial
($\sim50\%$).  We call these dark or obscured stars DS for short.
However, the fact that S0's and ellipticals are dynamically evolved
systems at the end of the gas-to-star transformation process, but with
still $\sim 40\%$ of dark matter, hints toward the presence of
non-gaseous components such as DS.  If we assume that a population of
DS pre-exists to a galaxy (the same r\^ole would be played by CDM
particles), its maximal fraction is determined by the final S0 stage,
$\sim40\%$.  In previous stages the difference must be gas to make
future stars.  Less speculatively, as shown in Fig.~1, if instead we
assume that DS's form proportional to the stars as galaxies evolve
(for example just from the increasing fraction of dust), the amount of
dark gas required is correspondingly increased.  For the Milky Way (an
SBbc) the fraction of DS's would be $<40\%$ in the first case, and
$<20\%$ in the second case.  

\section{Conclusions}
Galactic dynamics forces us to see spirals as structures with possible
``bursts'' of dynamical evolution involving short time-scales.  Taking
into account today's observational and theoretical constraints, the
only possible sense of evolution is from Sm to S0.  During evolution
everything happens as if dark matter in galaxies is transformed into
stars, that is dark gas is required.

A possible solution to the dark matter problem in galaxies is that gas
in outer HI disks clumps along a fractal structure down to solar
system sizes, as explained elsewhere in this volume. The smallest
clumps are then very dense and cold which makes them presently hard to
detect.  A sizable amount of dark (or dust obscured) stars can also be
argued just from the fact that evolved galaxies (Sa-S0's) still
contain about 40\% of dark matter.


\begin{thebibliography}{}

\bibitem{}{}{} 
Athanassoula E., Bienaym\'e O., Martinet L., Pfenniger D. 1983, {\it
A\&A} {\bf 127}, 349

\bibitem{}{}{}
Barnes J.E. 1992, {\it ApJ} {\bf 393}, 484

\bibitem{}{}{} 
Bothun G.D., Shombert J.M., Impey C.D., Schneider S.E. 1990, {\it ApJ}
{\bf 360}, 427

\bibitem{}{}{}
Broeils A. 1992, {\it Dark and visible matter in spiral galaxies}, 
PhD Thesis, U. Groningen

\bibitem{}{}{}
Butcher H., Oemler A. 1978, {\it ApJ} {\bf 219}, 18

\bibitem{}{}{}
Combes F., Debbasch F., Friedli D., Pfenniger D. 1990, {\it A\&A} {\bf
233}, 82

\bibitem{}{}{}
Combes F., Sanders R.H. 1981, {\it A\&A} {\bf 96}, 164

\bibitem{}{}{}
Contopoulos G. 1980, {\it A\&A} {\bf 81}, 198

\bibitem{}{}{} 
Davies J.I., Phillips S., Boyce P.J., Disney M.J. 1993, {\it MNRAS}
{\bf 260}, 491

\bibitem{}{}{}
de Vaucouleurs G. 1959, in {\it Handbuch der Physik LIII, Astrophysik
IV: Sternsysteme}, S. Fl\"ugge (ed.), Springer-Verlag, Berlin, 275

\bibitem{}{}{}
Dressler A. 1980, {\it ApJ} {\bf 236}, 351

\bibitem{}{}{}
Eggen O.J., Lynden-Bell D., Sandage A.R. 1962, {\it ApJ} {\bf 136},
748 (ELS)

\bibitem{}{}{}
Friedli D., Benz W. 1993, {\it A\&A} {\bf 268}, 65, and 
1995, {\it A\&A}, {\bf 301}, 649

\bibitem{}{}{}
Friedli D., Martinet L. 1993, {\it A\&A} {\bf 277}, 27

\bibitem{}{}{}
Hasan H., Norman C. 1990, {\it ApJ} {\bf 361}, 69.

\bibitem{}{}{}
Hohl F., Hockney R.W. 1969, {\it J.~Comput.~Phys.} {\bf 4}, 306

\bibitem{}{}{}
Kennicutt R.C. 1989, {\it ApJ} {\bf 344}, 685

\bibitem{}{}{} 
Kormendy J. 1982, in {\it Morphology and Dynamics of Galaxies}, 12th
Advanced Course Swiss Soc. Astr. Astrophys., Martinet L., Mayor
M. (eds.), Geneva Observ., 115

\bibitem{}{}{}
Larson R.B. 1981, {\it MNRAS}  {\bf 194}, 809

\bibitem{}{}{}
Larson R.B., Tinsley B.M., Caldwell C.N. 1980, {\it ApJ} {\bf 237}, 692

\bibitem{}{}{}
Martinet L. 1995, {\it Fundamental of Cosmic Physics} {\bf 15}, 341


\bibitem{}{}{}
Miller R.H., Prendergast K.H. 1968, {\it ApJ} {\bf 151}, 699

\bibitem{}{}{}
Pfenniger D. 1984, {\it A\&A} {\bf 134}, 373 

\bibitem{}{}{}
Pfenniger D. 1985, {\it A\&A} {\bf 150}, 112 

\bibitem{}{}{} 
Pfenniger D. 1991a, in {\it Dynamics of Disc Galaxies}, B.~Sundelius
(ed.), G\"oteborg U., 191

\bibitem{}{}{} 
Pfenniger D. 1991b, in {\it Dynamics of Disc Galaxies}, B.~Sundelius
(ed.), G\"oteborg U., 389

\bibitem{}{}{}
Pfenniger D. 1996, in {\it Third Paris Cosmology
Colloquium}, H.J. de Vega, N. S\'anchez (eds.), World Scientific,
Singapore, in press

\bibitem{}{}{}
Pfenniger D., Combes F. 1994, {\it A\&A} {\bf 285}, 94

\bibitem{}{}{}
Pfenniger D., Combes F., Martinet L. 1994, {\it A\&A} {\bf 285}, 79

\bibitem{}{}{}
Pfenniger D., Friedli D. 1991, {\it A\&A} {\bf 252}, 75

\bibitem{}{}{}
Pfenniger D., Norman C.A. 1990, {\it ApJ} {\bf 363}, 391

\bibitem{}{}{}
Quirk W.J. 1972, {\it ApJ} {\bf 176}, L9

\bibitem{}{}{}
Raha N., Sellwood J.A., James R.A., Kahn F.D. 1991, {\it Nat.} {\bf
352}, 411

\bibitem{}{}{}
Rakos K.D., Schombert J.M. 1995, {\it ApJ} {\bf 439}, 47

\bibitem{}{}{} 
Roberts M.S., Haynes M.P. 1994, {\it ARAA} {\bf 32}, 115

\bibitem{}{}{}
Safronov V.S. 1960, {\it Annales d'Astrophysique} {\bf 23}, 979

\bibitem{}{}{}
Schweizer F. 1993, in {\it Physics of Nearby Galaxies, Nature or
Nurture?}, T.X. Thuan, C. Balkowski, J.T.T. Van (eds.), Editions
Fronti\`eres, Gif-sur-Yvette, 283

\bibitem{}{}{}
Searle L., Zinn R. 1978 {\it ApJ} {\bf 225}, 357

\bibitem{}{}{}
Toomre A. 1964, {\it ApJ} {\bf 139}, 1217

\bibitem{}{}{}
Toomre A. 1981, in {\it The Structure and Evolution of Normal
Galaxies}, S.M. Fall, D. Lynden-Bell (eds.), Cambridge Univ.~Press, 111

\bibitem{}{}{}
Toomre A., Toomre J. 1972, {\it ApJ} {\bf 178}, 623

\bibitem{}{}{}
T\'oth G., Ostriker J.P. 1992, {\it ApJ} {\bf 389}, 5

\bibitem{}{}{} 
White S.D.M. 1994, {\it Formation and Evolution of Galaxies}: Les
Houches Lectures, preprint astro-ph/9410043

\bibitem{}{}{}
Whitmore B.C. 1993, in {\it Physics of Nearby Galaxies, Nature or
Nurture?}, T.X. Thuan, C. Balkowski, J.T.T. Van (eds.), Editions
Fronti\`eres, Gif-sur-Yvette, 425

\bibitem{}{}{}
Young J.S., Knezek P.M. 1989, {\it ApJ} {\bf 347}, L55 

\bibitem{}{}{}
Zaritsky D., Kennicutt R.C., Huchra J.P. 1994, {\it ApJ} {\bf 420}, 87

\end{thebibliography}
\end{document}